\documentstyle[twocolumn,prb,aps]{revtex}
\begin{document}

% this places title+abstract over two column widths
\twocolumn[\hsize\textwidth\columnwidth\hsize\csname @twocolumnfalse\endcsname 

\date{June 20, 1996}
%\titlepage
\draft
\title{\bf Edge Excitations of the $\nu$ = $\case{2}{3}$
Spin-Singlet Quantum Hall State}
\author{J.\ E.\ Moore~\cite{whereamI}\cite{email} and
F.\ D.\ M.\ Haldane }
\address{Department of Physics \\
Princeton University \\
Princeton, N.J. 08544-0708}
\maketitle
\begin{abstract}
The spectrum of edge excitations is derived for the spin-unpolarized
$\nu = 2$ and $\nu = 2/3$ FQHE.  Numerical diagonalization of a system
of six electrons on a disc confirms that the edge $\nu = 2/3$ spin-singlet
FQHE state consists of oppositely directed spin and charge branches
on the same physical edge.  The highly correlated $\nu = 2/3$ singlet edge
is shown to have the same spin branch as the $\nu = 2$ singlet edge,
providing evidence that the same $SU(2)_{k = 1}$ Kac-Moody algebra describes
all unmixed spin branches.  The spin and charge branches of the singlet state
at $\nu = 2/3$ are less coupled than the two branches
of the spin-polarized state at the same filling factor, suggesting that
the conductance along an edge may increase sharply across the
polarized-unpolarized transition.
\end{abstract}
\pacs{PACS numbers: 72.20.My 73.20.Dx 73.40.Hm}

% ] marks end of single wide column text region of \twocolumn[...text...]
]

\narrowtext

\section{Introduction}
The possibility of spin-singlet ground states in the fractional quantum
Hall effect
(FQHE) has
received considerable experimental and theoretical attention.  The
ground state at filling factor $\nu = 2/3$ is found experimentally~\cite
{clark,eisenstein} and numerically~\cite{chakra,maksym,jkj,macdon} to
be spin-unpolarized for low magnetic field but spin-polarized
for high field.
% specific nature of edge measurements
Halperin~\cite{halperin} first suggested this possibility on the
grounds that the ratio of Zeeman energy to cyclotron energy is
$\sim {1 \over 60}$ for GaAs in a weak magnetic field, so that electron-electron
interactions may override the tendency of electron spins to align with
the magnetic field while still being too weak to push electrons into higher
Landau levels.
The spin-polarized state at $\nu = 2/3$ is now well understood as the
particle-hole
transform of the Laughlin $\nu = 1/3$ state~\cite{laughlin}, and
its edge has been
shown numerically to be a combination of two spatially separated sub-edges
~\cite{johnson,wenrev}: an inner edge where the density of holes drops
from $1/3$ to $0$, and an outer edge where the density of electrons drops
from $1$ to $0$.

This paper studies singlet FQHE edges and in particular the edge of the
$\nu = 2/3$ unpolarized state, which also
contains two oppositely directed branches but does not break down into
separate edges.  The $\nu = 2/3$ unpolarized state carries oppositely chiral
spin and charge branches of excitations on the same physical edge.
Its spin excitations are described by the $SU(2)_{k=1}$ Kac-Moody
algebra and thus have a different structure from previously studied
$U(1)_{k=1}$ 
charge excitations.  The $\nu = 2/3$ edge inherits its structure
from the properties of the bulk FQHE liquid, and part of this structure
is shown to arise as a nontrivial prediction of the Chern-Simons effective
theory of the FQHE~\cite{chern}.

The excitation spectra of FQHE edges are of interest for a number of reasons. 
Edge excitations remain gapless in the thermodynamic limit, unlike
excitations in the bulk fluid, and thus may determine the specific
heat.  They can provide a nonzero conductivity
between two contacts connected by an edge.
Finally, FQHE edges are more
accessible experimentally than the bulk liquids and may
provide the only practical probe of the rich structure of FQHE liquids.

The same $SU(2)$ algebra is expected to apply equally well to all spin
branches which are not mixed with other modes.  The spin and charge branches
of the $\nu = 2/3$ edge are found numerically to be
independent to 1 part in $10^4$.  Hence the spectrum of the spin branch
should be given by the same algebra describing spin excitations at
$\nu = 2$.  We can derive the pattern of excitations for a simple
singlet system, the $\nu = 2$ state,
and expect the same pattern to describe the spin branch of the
$\nu = 2/3$ highly correlated system.
% This technique yields the
% spectrum while avoiding the
% full derivation of the $SU(2)$ Kac-Moody algebra of operators at the 
% edge.
(How the $SU(2)$ algebra arises from the Chern-Simons bulk
theory of the $\nu = 2/3$ state is discussed briefly below and
in more detail in McDonald and Haldane~\cite{mcd})
The stability of the $SU(2)$
Kac-Moody algebra, not present in the $U(1)$ case, is mathematically
a consequence of the 
additional non-Abelian structure of the $SU(2)$ algebra.  In physical
terms, strong electron-electron interactions might be expected to modify the
effective charge of an electron but not its spin.

The $\nu = 2/3$ unpolarized edge has previously attracted theoretical
interest as
the simplest spin-singlet FQHE edge, but often at a rather high
level of abstraction.  We carried out a numerical
diagonalization of a system of six interacting
electrons at $\nu = 2/3$ and observed the predicted pattern of
excitations.
The disc geometry was used as the simplest geometry
possessing an edge.  The numerical results indirectly confirm the
topological properties ($K$-matrix and shift) of the wavefunction
proposed by Wu, Dev, and Jain~\cite{jkj}.   The $\nu = 2/3$ edge
is seen to consist of two oppositely directed branches,
verifying a nontrivial
prediction of the chiral Luttinger liquid theory of FQHE edges.
The same prediction is shown to arise from the quantization of
total angular momentum of electrons on a disc.
The differences at the edge between polarized and unpolarized states
at $\nu = 2/3$ may yield a useful experimental probe of the
polarized-unpolarized
transition.

\section{Theory of the singlet edge}

The edge excitations of FQHE liquids have a rich
topological structure which is closely related to the Chern-Simons
effective theory of the FQHE.  The Hilbert space of excitations of
the $\nu = 2/3$ unpolarized state is a direct product of oppositely
directed spin and charge sectors~\cite{wen,balatsky}.  The $U(1)$ Kac-Moody
algebra of the charge sector and its associated spectrum are now well
understood in the FQHE context, but the $SU(2)$
algebra of
the spin sector gives a more complicated excitation spectrum related to a
different conformal field theory.  For concreteness, we examine
the $\nu = 2$ and
$\nu = 2/3$ singlet edges and obtain
testable predictions about their excitation spectra.

The spectrum for a single branch of edge excitations on a droplet
of $N$ electrons has been derived from quantization of
hydrodynamic perturbations on the edge of a charged fluid~\cite{wen}
and from the polynomial representation of lowest-Landau-level
wavefunctions~\cite{hald}.  This is the spectrum of excitations
for the Laughlin-type states at $\nu = 1/m$, $m$ odd.  The result is
\begin{equation}
E = {M v \over L } = {v \hbar \sum_{m = 0}^\infty
{m c_m^{\dagger} c_m} \over L} \label{spect}
\end{equation}
where $v$ is the branch velocity, $L = 2 \pi R$ the droplet circumference,
$M$ total angular
momentum, and the $c_m$ are
independent bosonic modes of angular momentum $m \hbar$.

The simplest singlet state is the $\nu = 2$ state consisting of filled
spin-up and spin-down lowest Landau levels.  The excitations of this
state can be explicitly calculated within the lowest Landau level
using the rotationally symmetric single-electron wavefunctions
\begin{equation}
\psi_m = {e^{i \theta m} r^m e^{-r^2 / 4 \ell^2}
\over \ell^{m + 1} \sqrt{2^{m + 1} \pi m}} = {z^m e^{-|z|^2 / 4 \ell^2}
\over \ell^{m + 1} \sqrt{2^{m + 1} \pi m}}.
\label{orbs}
\end{equation}
The ground state at $\nu = 2$ of $N = 2k$ electrons confined by some
radial potential has all orbitals $m = 0,\ldots,k - 1$ filled with both up
and down electrons.  This state has total angular momentum $M_0 = k (k-1)$
in units with $\hbar = 1$.  Its excitations
within the lowest Landau level consist of (i) some combination of excitations
of the up spins, with corresponding creation and annihilation operators
$b_i,b^\dagger_i$, (ii) some combination of
excitations of the down spins, denoted by $c_i,c^\dagger_i$, and (iii)
some (positive or negative) number $S_z$ of up spins being converted to down
spins.  Clearly $S_z$ is the z-component of spin of the resulting state,
and all excited states will be eigenstates of $S_z$ because the
Hamiltonian is spin-symmetric.
The total angular momentum of an excited state is, using (\ref{spect}),
\begin{equation}
M = M_0 + {S_z}^2 + \sum_l{l b^\dagger_l b_l} + \sum_m{m c^\dagger_m c_m}.
\label{mom}
\end{equation}

The excitations must still be classified into spin and charge branches,
which may have different velocities.  Some excitations are purely spin
excitations, some are purely charge excitations, and some are mixed.
The pure charge excitations are the hydrodynamic modes of a charged
droplet and thus have the spectrum (\ref{spect}).  Hence one of the
two sums over bosonic modes in $(\ref{mom})$ corresponds to the charge
branch.  The other sum must correspond to the spin branch, so that the
total edge Hamiltonian is
\begin{equation}
H = {v_s \over L} ({S_z}^2 + \sum_l{l b^\dagger_l b_l}) +
{v_c \over L} {\sum_m{m c^\dagger_m c_m}}.
\label{s2ham}
\end{equation}
Here $v_s$ and $v_c$ are the (positive) velocities of the spin and
charge branches.
Pure spin (charge) excitations are given by states which are annihilated
by all $c_m$ ($b_m$).  The resulting structure of pure spin excitations
is given in Table \ref{tone}.  If the system is exactly spin-symmetric
then the excited states are unified into exact $SU(2)$ multiplets.

The spin branch in (\ref{s2ham})
is closely related to the Virasoro operator $L_0$ of a conformal field
theory (CFT) of free bosons in $1 + 1$
dimensions.  The spin branch corresponds to a different CFT than the
charged excitations of a $\nu = 1/m$ state.  The charged excitations of
a single charge branch with $\Delta N \geq 0$ have angular momentum
\begin{equation}
L = {(\Delta N)^2 \over 2} + \sum_{m = 0}^\infty{c_m^\dagger c_m}.
\label{charge}
\end{equation}
Free bosonic CFT's in $1 + 1$ dimensions contain in the generator $L_0$ a term
proportional to $N^2$ with coefficient depending on the boson radius.
The single-branch spectrum (\ref{charge}) is given by
the CFT with boson radius of compactification $1$, while the boson radius
is $1/2$ for the spin branch if $S_z$ is taken to correspond to winding
number $N$.

The spectrum (\ref{s2ham}) describes the edge excitations of the
$\nu = 2$ polarized state, which has two branches moving in the same
direction.  The charge and spin branches have
oppositely directed velocities
for $\nu = 2/3$, however.  This is a nontrivial prediction of the
Chern-Simons effective theory of the FQHE~\cite{chern}, which yields
a ``chiral Luttinger liquid'' theory on the edge~\cite{wen}.  In this
theory the $\nu = 2/3$ unpolarized state is associated with the K-matrix
$({1 \atop 2}
{2 \atop 1})$ (up to $SL(2,Z)$ equivalence).  The edge consists of two
oppositely directed branches
because this
matrix has one positive and one negative eigenvalue.
The $\nu = 2/3$ unpolarized edge has the same charge
and spin branches as derived above for independent electrons;
the sole difference is that the branches are oppositely directed.
The charge branch, corresponding to hydrodynamic perturbations, has
excitations of positive angular momentum, and the spin branch has
excitations of negative angular momentum.

The fact that the edge branches
propagate in opposite directions is not clearly evident in the
explicit wavefunction of
the ground state~\cite{jkj}, nor can it be derived from a hydrodynamic
approach because the $\nu = 2$ and $\nu = 2/3$ state would seem to have
the same hydrodynamics.  The reversed direction of the spin branch
is observed in numerical studies, confirming the prediction of the
effective theory of the FQHE and reflecting the deep connection
between the bulk FQHE liquid and edge excitations.  The Halperin-Laughlin
multicomponent
wavefunctions all have positive definite $K$-matrices and hence cannot
describe the observed $\nu = 2/3$ state.~\cite{mcd}
The $\nu = 2/3$
FQHE edge
is particularly significant for theory because
it is described by a non-Fermi-liquid theory at constant N.  The
$U(1)$
Kac-Moody algebra which describes a single edge branch exhibits a
non-Fermi-liquid structure only with variable $N$~\cite{wen}.

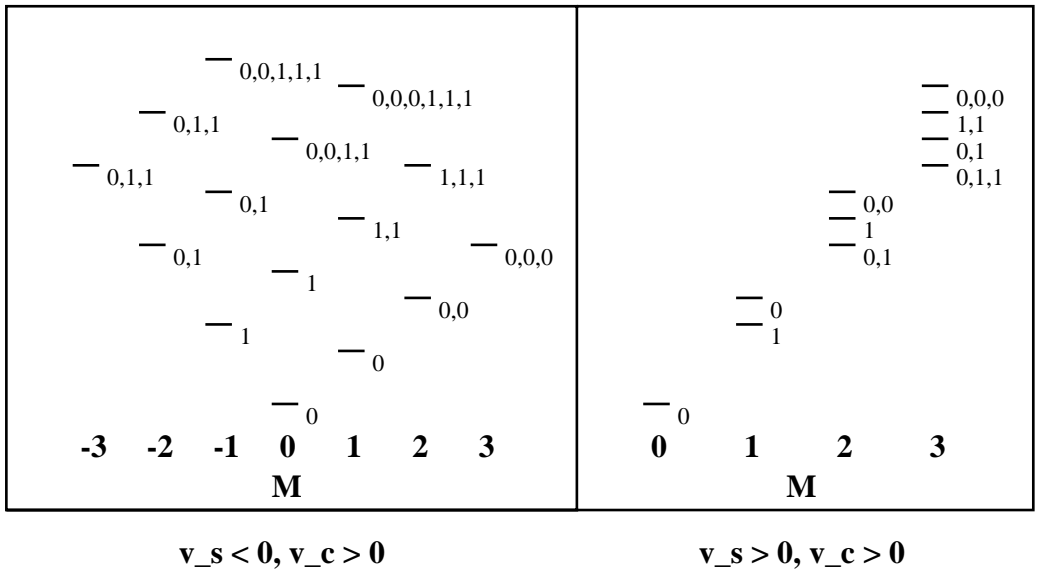
\begin{figure}
\caption{Predicted spectrum of edge excitations for two-component
edge states possessing an $SU(2)$ symmetry.  The states have been combined
into $SU(2)$ multiplets, so that $(0,1)$ denotes a fourfold degeneracy
of one $S^2 = 0$ and three $S^2 = 2 \hbar$ states.}
\label{figzero}
\end{figure}

Fig.~\ref{figzero} shows ideal spectra for the $\nu = 2/3$ (opposite
velocities)
and $\nu = 2$ (parallel velocities) edges, where the velocities
have been drawn
with magnitudes as is generically the case.  The states
have been combined into exact $SU(2)$ multiplets.

Some insight into the nature of the spin branch in the $\nu = 2$ electrons
can be gained by considering an ideal system of electrons which interact
only through Pauli exclusion.  Suppose that there is one orbital at
each value of the angular momentum $m$, which can be filled by at most
one spin-up electron and one spin-down electron, and that these orbitals
have energies $E(m)$.  If the ground state is a singlet, its energy must
be increased by the operation of transferring one spin-up electron to
spin-down (and vice versa).  Since energy is a smooth function of
occupancies in the thermodynamic limit, there will
be a term in the Hamiltonian containing $(N_\uparrow - N_\downarrow)^2
= {S_z}^2$.  Excitations which
do not change $S_z$ will have energies (and hence velocity) proportional
to $dE / dm$.  The prediction derived above for the spectrum amounts
to a fixed ratio between the $S_z$ term and the constant-spin excitation
term.  The ratio is the same value as obtained for an exactly linear
dispersion relation $E(m)$, i.e., the same as the requirement
used above that the energies of edge excitations be linear in the angular
momentum.

\section{Connection to the chiral Luttinger liquid theory}

These results on the structure of singlet edges can be placed in the
context of field theories of the FQHE edge.  The simplest edge is
the $\nu = 1/m$ edge, which possesses only a single branch of
excitations.  This edge is described as
a ``chiral Luttinger liquid'' with $1 + 1$-dimensional action
\begin{equation}
S = -{\hbar \over 4 \pi} \oint{\partial_x \phi \partial_t
	\phi + v (\partial_x \phi)^2 \, dx}.
\end{equation}
The edge state theory predicts that the charge-neutral edge excitations of an
abelian quantum Hall state containing $N$ species of quasiparticles
form a representation
of $N$ $U(1)$ Kac-Moody algebras.  This result can be obtained from
quantization of the above edge action for $\nu = 1/m$
states or from the multispecies chiral Luttinger liquid picture
of more complicated
states.  The condition for a general multispecies edge
to admit an $SU(2)$ symmetry can also be derived from the
theory.~\cite{mcd}.  For a two-quasiparticle state such as the $\nu = 2$ or
$\nu = 2/3$ unpolarized states, there are two families of edge
operators satisfying the $U(1)$ Kac-Moody algebra commutation relations:

\begin{equation}
[\rho_{k^\prime},\rho_k] = v_1 k \delta_{k + k^\prime} 
\end{equation}
\begin{equation}
[\tilde \rho_{k^\prime},\tilde \rho_k] = v_2 k \delta_{k + k^\prime}.
\label{commrels}
\end{equation}
These can be combined to give raising and lowering operators for
two families of harmonic oscillators.
The angular momentum and Hamiltonian of the system are
\begin{equation}
L = L_0 + \sum_{k=1}^\infty{{v_1 \over |v_1|} \rho_k \rho_{-k}
	+ {v_2 \over |v_2|} {\tilde \rho}_k {\tilde \rho}_{-k}}
\end{equation}
\begin{equation}
H = \sum_{k = 1}^\infty{{|v_1| \over L} \rho_k \rho_{-k}
+ {|v_2| \over L} {\tilde \rho}_k {\tilde \rho}_{-k}}.
\end{equation}

At this point we must add operators to the system reflecting its
$SU(2)$ symmetry.  These operators cannot be implicit in the
$K$-matrix describing the bulk state because the long-distance
behavior reflected in the effective theory does not determine the
symmetries.  As a concrete example, the $\nu = 2/3$ polarized
and unpolarized states have the same $K$-matrix but different
symmetry properties.

The spin operators commute with the $U(1)$
excitation operators and appear in the angular momentum and
Hamiltonian only as terms proportional to ${S_z}^2$, as derived above.
Therefore $S^2$ and $S_z$ are good quantum numbers for edge states.
Taking $v_1, \rho_k$ to give the spin branch and $v_2,
{\tilde \rho}_k$ the charge branch,
\begin{equation}
[S_x, \rho_k]= [S_y, \rho_k] = [S_z, \rho_k] = 0
\end{equation}
\begin{equation}
[S_x, {\tilde \rho}_k]= [S_y, {\tilde \rho}_k] = [S_z, {\tilde \rho}_k] = 0
\end{equation}
\begin{equation} 
L = L_0 + \sum_{k=1}^\infty{{v_1 \over |v_1|} ({S_z}^2 + \rho_k \rho_{-k})
	+ {v_2 \over |v_2|} {\tilde \rho}_k {\tilde \rho}_{-k}}
\end{equation}
\begin{equation}
H = \sum_{k = 1}^\infty{{|v_1| \over L} ({S_z}^2 + \rho_k \rho_{-k})
+ {|v_2| \over L} {\tilde \rho}_k {\tilde \rho}_{-k}}.
\end{equation}

The edge operators in the chiral Luttinger liquid picture
of a singlet state form a representation of the $SU(2)$ Kac-Moody algebra.
At constant $N$ but variable spin, the excitation spectrum differs from the
ordinary $U(1)$ spectrum because of the ${S_z}^2$ term (Table \ref{tone}).
The states in
Table \ref{tone} must be reunified into $SU(2)$ multiplets because of the $SU(2)$ spin
symmetry of the system.  Thus one of the $|\Delta M| =
2$ $S_z = 0$ states is exactly degenerate with the $S_z = \pm 1$ states,
for example.

\section{Results from numerical diagonalization}
\label{snumer}

Now we examine the numerical spectrum obtained by diagonalizing the
Hamiltonian for a disc of six interacting
electrons in the lowest Landau level:
\begin{equation}
H = {1 \over 2} \sum_{i,j,k,l} {V_{ijkl}
a_j^\dagger a_i^\dagger a_k a_l} + \sum_i{U_i a_i a_i^\dagger}.
\label{hamil}
\end{equation}
The electrons in our simulation are confined by a harmonic potential $U(r) =
\alpha r^2$ where $\alpha$ is chosen to give central density $\nu / 2 \pi
\ell^2,$ where $\ell = \sqrt{\hbar c / e B}$ is the magnetic length.  Any
radial potential retains the conservation of total angular
momentum, so that each angular momentum subspace can be diagonalized
independently.  The
qualitative nature of the edge excitations is independent of the choice
of $U(r)$ as long as the edge separates $\nu = 2/3$ from $\nu = 0$ with no
intermediate Hall liquid.

We study the $\nu = 2/3$ state rather than the $\nu = 2$ unpolarized
state because all excitations of the $\nu = 2$ state are edge excitations,
because all orbitals inside the edge are fully occupied for states not
too excited from the
ground state.  In the same way all low-lying polarized excitations of
the $\nu = 1$ state are edge excitations.  The number of excitations
at given angular momentum can be exactly calculated
for the $\nu = 2$ unpolarized state (as done above) and the $\nu = 1$
polarized state.  Numerical calculation is not especially illuminating
for these states because the numbers of excitations of given angular momentum
and spin are fixed simply by state counting; the only information gained by
a numerical calculation is the velocity of the branches, which is not
expected to be universal anyway.  Calculations of $\nu = 1/3,2/3,2/5$
etc. are useful precisely because not every excited state is to be
interpreted as an edge state.  If the low-lying excitations, which
are much smaller in number than the set of all excitations, follow
the theoretically predicted pattern for one of these states, it is
strong confirmation of the theory.

The effect of the interaction potential $V$ in the lowest Landau level is
completely determined by the pseudopotentials~\cite{pseudo}
\begin{equation}
V_m = {\int_0^\infty{dr\,r^{2 m + 1} V(r) \exp(- r^2 / 4 \ell^2)}
\over \int_0^\infty{dr\,r^{2 m + 1} \exp(- r^2 / 4
\ell^2)}}.
\label{pseudeq}
\end{equation}
The highly correlated structure of FQHE wavefunctions is typically destroyed
by strong long-range pseudopotentials.  This is manifested numerically by the
disappearance of the bulk excitation gap characteristic of the FQHE.  Such an
excitation gap is useful for edge studies because (gapless) edge excitations
are clearly distinguishable in the spectrum from bulk excitations.
The low-lying excited states can be verified to be edge excitations by
checking that they differ from the ground state only near the edge.

The rotational symmetry of the disc
geometry makes convenient the single-electron basis (\ref{orbs}).
% \begin{equation}
% \psi_m = {e^{i \theta m} r^m e^{-r^2 / 4 \ell^2}
% \over \ell^{m + 1} \sqrt{2^{m + 1} \pi m}} = {z^m e^{-|z|^2 / 4 \ell^2}
% \over \ell^{m + 1} \sqrt{2^{m + 1} \pi m}}.
% \label{orbitals}
% \end{equation}
The basis states of the diagonalization are Slater determinants formed from
these orbitals.  The Hilbert space is made finite by neglecting those
$\psi_m$ which are
negligible
in the vicinity of the droplet of electrons.

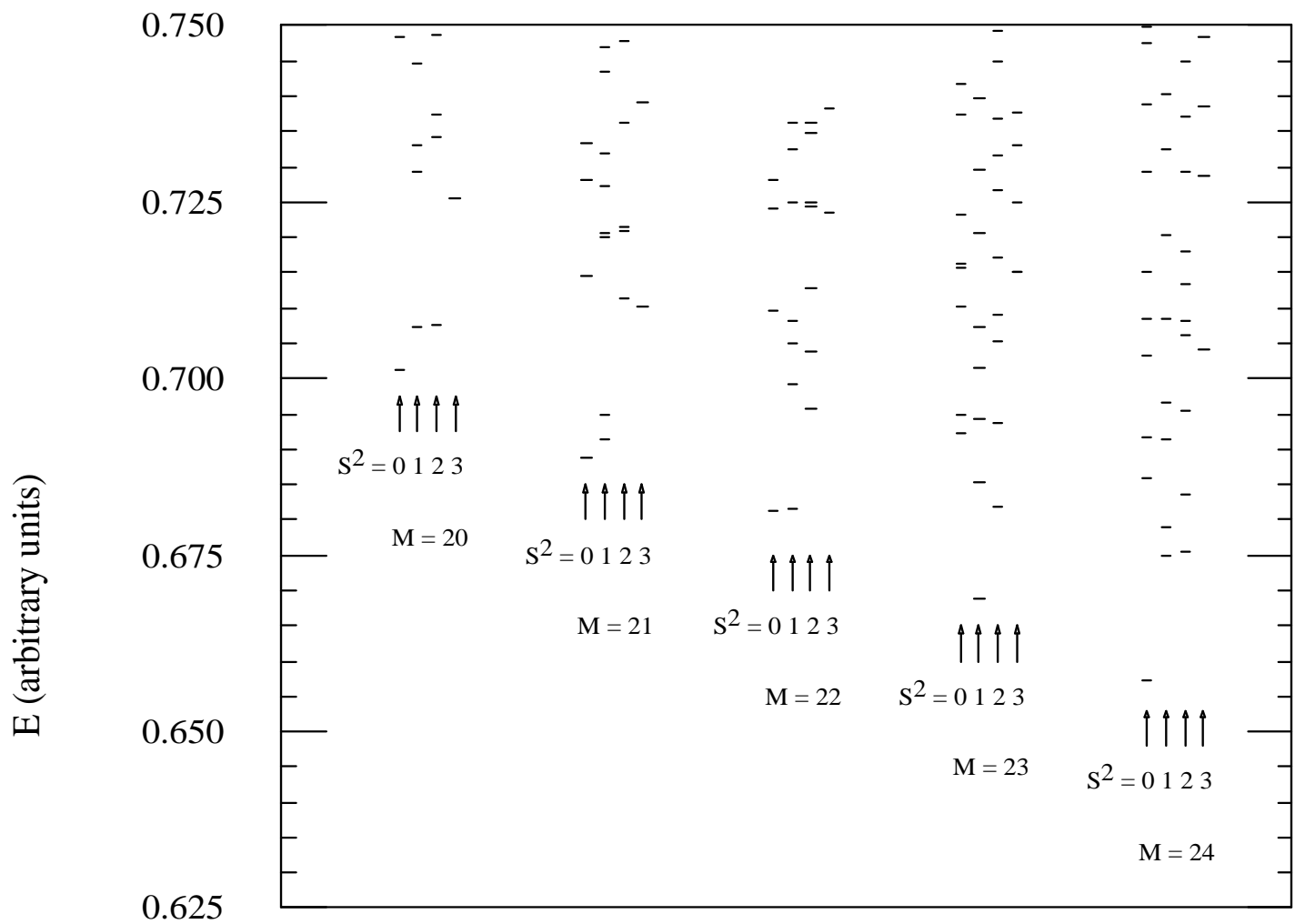
\begin{figure}
\caption{Energy levels from numerical diagonalization of six interacting
electrons in two dimensions.  The electrons interact through a screened
Coulomb potential and are confined by a (radially symmetric)
harmonic potential.}
\label{figtwo}
\end{figure}

The observed pattern of the spectrum near the ground state is remarkably
similar for a variety of short-range interactions.  Long-range interactions
such as the unscreened Coulomb interaction $V_m = \Gamma((n + 1) / 2) /
2 \Gamma(n/2 + 1)$ destroy the Kac-Moody $SU(2)_{k = 1}$ pattern of edge
excitations but also destroy the FQHE.  None of the interactions tested
destroys the edge excitations without eliminating the excitation gap
characteristic of the FQHE.  Fig. \ref{figtwo} shows the lowest-energy states 
for the screened Coulomb potential
\begin{equation}
V(r) = {e^2 \exp(- r / \ell) \over r^2}
\label{cutoffp}
\end{equation}
The pattern begins to disappear when the screening length in (\ref{cutoffp})
is
increased to $2
\ell$.

Fig. \ref{figtwo} shows the excitations of negative angular momentum
relative to the ground state.
Single-particle density plots confirm that the low-lying states differ
from the ground state only in the vicinity of the edge.  Each angular
momentum subspace contains of order $10^3$ states, of which only the
lowest few are shown.  There is a
well-separated ground state at $M = 24 =
{1 \over 2} N N_\phi$.  This is the ground state angular momentum for the
wavefunction of Wu et al.~\cite{jkj}, which in the disc geometry is
\begin{eqnarray}
\Psi = e^{-\sum_i{z_i^2 \over 4 \ell^2}}&(\prod_{i<j}(\partial_{z_i} - \partial_{z_j})^{\delta_{\sigma_i
\sigma_j}} e^{{i \pi \over 2} {\rm sgn}(\sigma_i - \sigma_j)})
\nonumber \\
&\times(\prod_{i < j}
(z_i - z_j)^2).\label{wavefn}
\end{eqnarray}
Hence $N \phi = {3 N \over 2} - 1$ is the number of flux quanta,
confirming that 1 is the shift of the
singlet state on the sphere.

The Hamiltonian in the
absence of a Zeeman term contains no spin terms, so that the calculated
spectrum contains exactly degenerate $SU(2)$ multiplets.  For example, each
$S^2 = 1$ energy in Fig. \ref{figtwo} represents three states with $S_z = -1,0,1$. 
The basis for the diagonalization consists of all six-particle
Slater determinants formed from spin-up and spin-down versions of the $\psi_m$
in (\ref{orbs}).  The basis states are eigenstates of $S_z$ but not $S^2$,
but $S^2$ is easily calculated for the energy eigenstates
using $S^2 = S_+ S_- + S_- S_+ + {S_z}^2.$

The calculated spectrum (Fig. \ref{figtwo}) shows some splitting in the
degeneracies of the pattern (\ref{s2ham}), presumably caused by finite-size
effects.  But the features of the numerical
spectrum are well described by the pure spin branch spectrum given in
Table \ref{ttwo}.
The pure spin excitations up to $M = 20$
are clearly
separated from the clutter of bulk excitations.  $M = 20$
($\Delta M = - 4$) is the first subspace found to
have a low-lying state with $S^2 = 2$, matching the theoretical prediction.
This pattern is strong confirmation for the theory derived above partly
because the na{\"\i}ve addition of excitations gives a different
pattern: for example, the sum of two $\Delta M = - 1$ excitations with $S^2 = 1$
would give $S^2 = 0,1,2$ at $M = 22$ rather than the observed $S^2 = 0,1$.

The energies of spin and charge excitations are found to add to
one part in $10^4$ of the excitation energy.  This allows the identification
of some states in the bulk as mixed edge excitations because their energies
are almost exactly the values expected from energy addition: the error
is typically about $1\%$ of the typical inter-level distance.
The
implications of this result can be understood by imagining that the
Hamiltonian for the
system consists of an ideal Hamiltonian giving the exact
structure plus a perturbation $V$.  The additivity of energies means that
$V$ couples oppositely directed modes weakly, in contrast to the numerical
results~\cite{johnson,wenrev} for the $\nu = 2/3$ polarized edge. 
Since the perturbation $V$
from finite-size effects is essentially arbitrary, we can expect the
unpolarized edge to have reduced backscattering (relative to the
polarized edge)
for a generic perturbation.  This should be manifested in real systems as
a decrease in the finite two-point resistance along an edge as the
polarized-unpolarized transition occurs.  The reduced backscattering can be
understood simply: almost all excitations in the spin sector have
non-zero $S^2$, so that spin-independent perturbations will cause no
backscattering.

\section{Implications of quantization of total angular momentum}

The total angular momentum of a system of $N$ electrons is of
course quantized in units of $\hbar$.  This simple statement applied
to singlet states yields
a relation between the direction of the spin branch at the edge
and the topological shift of the bulk wavefunction.  The results
from this calculation match the numerical results obtained above
for the $\nu = 2/3$ state and also predict an excess energy for
systems containing an odd number of electrons.

The total angular momentum of the ground state of $N$ electrons
at filling factor $\nu$ is $M = {1 \over 2} N N_\phi$, where
\begin{equation}
N_\phi = {N \over \nu} - S
\end{equation}
is the number of flux quanta in the state and S is the ``shift'' of
the wavefunction on the sphere.  The angular momentum of the ground
state is thus
\begin{equation}
L_0 = {N^2  - \nu N S \over 2 \nu} \pm {S_z}^2
\end{equation}
with the positive sign corresponding to a spin branch of positive velocity,
i.e., of the same sign as the charge branch, and the negative sign
to an oppositely directed branch.

The states with $\nu = 2/m$ are the simplest singlet states.
For $N$ even, $S_z = 0$ in the singlet ground state and $L_0$ is integral
since $(N^2 m / 4) - (N S / 2)$ is integral for even $N$ for any integers
$m$ and $S$.  For $N$ odd the constraint that $L$ be integral requires
that $(m / 4) - (S / 2) \pm {S_z}^2$ be integral.  The lowest-energy
state for odd $N$ has ${S_z}^2 = 1/4$, so there are four possibilities:
\par\medskip
\hbox{\hskip 0.25 in
\vbox{
\noindent I. $S$ odd, $m = 4 k + 1$, spin velocity positive \par
\noindent II. $S$ \rm{odd,} $m = 4 k + 3$, spin velocity negative \par
\noindent III. $S$ \rm{even,} $m = 4 k + 1$, spin velocity positive \par
\noindent IV. $S$ \rm{even,} $m = 4 k + 3$, spin velocity negative.}\hfil}
\par
\medskip
\noindent
The singlet $\nu = 2$ state belongs to category I and the singlet
$\nu = 2/3$ state to category II, since both of these states have
shift $S = 1$.  If the $\nu = 2/5$ unpolarized state has $K$
and $S$ equal to their values in the hierarchy $\nu = 2/5$ polarized
state, as happens at $\nu = 2/3$, then a positively directed spin
branch exists at $\nu = 2/5$.

Another consequence is that states with an odd
number of electrons should be displaced in energy from states with
an even number of electrons.  The term $\hbar v_s {S_z}^2 / L$
in the excitation
Hamiltonian (\ref{s2ham}) should contribute an energy $\hbar v_s / 4 L$
to the doubly degenerate $S_z = \pm 1/2$ ground multiplet of odd numbers
of electrons.  Thus the ground state
energy $E(N)$ of $N$ electrons should satisfy
\begin{equation}
{E(2 k -1) + E(2 k + 1) \over 2} - E(2 k) = {\hbar v_s \over 4 L}
\end{equation}
while the right side of this equation would be $0$ for an ordinary
system with a well-defined chemical potential, i.e., $\mu = E^\prime(N)
\gg E^{(k)}(N), k > 1$.  This excess energy may also receive a
contribution from the $\Delta N = 1$ charged mode at the edge,
since the ground state of an odd number of particles is a
$\Delta N = 1$ excitation of an even-number ground state.
The energy of this charge excitation can be estimated by
writing $N = N_0 + \Delta N$ and interpreting the resulting term
\begin{equation}
L_{\rm charge} = {(\Delta N)^2 \over 2 \nu}
\end{equation}
as resulting from the charged mode.  The interpretation is correct
in the $\nu = 2$ case for a linearized Fermi surface, but it is unclear
to what degree the energy of the charged mode will be renormalized by
interactions.  This predicts a total energy
excess of
\begin{equation}
\Delta E = {\hbar v_c \over 2 \nu} + {\hbar v_s \over 4}.
\end{equation}
We have attempted to
see this effect numerically, but for small systems nonlinearities in
$E(N)$ from finite size swamp the desired effect.

\section{Conclusions}

The spin branch of a singlet FQHE edge has a complicated structure which is
difficult to explain from a purely hydrodynamical point of view.  The
$SU(2)$ symmetry of the singlet FQHE is likely to be the only exact
symmetry of accessible FQHE states, so the spin-branch spectrum described
here and the previously derived charge-branch spectrum may be the only
excitation patterns observable in experiments.

The numerical results obtained are strong evidence for the
$SU(2)_s \otimes U(1)_c$ description of the $\nu = 2/3$ singlet edge.  The
validity of the spectrum obtained from free fermions for
the highly correlated $\nu = 2/3$ state suggests that the same
pattern of singlet spin excitations describes all unmixed
spin branches, including the $\nu = 2$ spin branch.  The $\nu = 2/3$ FQHE
is experimentally quite robust, and
theoretical predictions about the edge excitations may be directly observable.

Edges of FQHE states inherit their structure from the bulk electron liquid.
Their transformation properties under gauge transformations of the
Chern-Simons effective theory of the bulk give rise to quantitative
connections between the edge spectra and bulk properties.  The $\nu = 2/3$
unpolarized state has an indefinite $K$-matrix in the effective theory
and hence has edge modes of both chiralities, as seen numerically.
One consequence of this is that the $\nu = 2/3$ singlet
state shows non-Fermi-liquid behavior even at constant $N$,
unlike $\nu = 1/m$ edges.

Some predictions of the effective theory
arise simply from the requirement of integral angular
momentum, which gives a relation between filling factor, shift,
and edge chirality satisfied by the $\nu = 2$ and $\nu = 2/3$ states
and possibly by all singlet states.  Another result is a prediction
that the ground state energy of an odd number of
electrons is raised by an energy $\hbar v_s / 4$ relative to the
energy of an even number of electrons, but we have not yet been able to
verify this conjecture numerically.

The description of the excitation spectrum as a direct product is more
accurate in the unpolarized edge than in the polarized edge,
suggesting that the unpolarized edge may have decreased backscattering and
increased two-probe conductance along an edge.  The measurement of edge
conductance may reveal interesting properties of the
$\nu = 2/3$ polarized-unpolarized transition.

\section{Acknowledgments}
	The authors would like to acknowledge useful conversations with
Kun Yang and J. Talstra.  J. E. M. was supported by a U. S. Fulbright
grant for part of the period during which this work was completed.
This work was supported in part by NSF DMR-9400362.

\begin{table}
\caption{Number of edge excitations for given values of $S_z$.
Excitations with the same $\Delta_M$ belong to the same Kac-Moody multiplet and
are expected to be approximately degenerate.  The states will form
exactly degenerate ordinary $SU(2)$ multiplets because of the
unbroken spin symmetry of the Hamiltonian.} \label{tone}
\end{table}

\begin{table}
\caption{Predicted $SU(2)$ multiplets for various values of
total angular momentum $M$ of six electrons at filling factor $\nu = 2/3$.
The ground state at $M$ = 24 supports spin excitations of negative angular
momentum in addition to charge excitations of positive angular momentum
which are not shown.} \label{ttwo}
\end{table}

\eject
\vfil
\centerline{\bf Table \ref{tone}}
\bigskip\bigskip
\hfil\hbox{\vrule\vbox{\offinterlineskip
\hrule
\vskip2pt\hrule
\halign {&\phantom{\vrule#}&\quad\hfil#\hfil\quad\cr
height5pt&&\multispan8\hfil&\cr
&\multispan9\hfil Number of modes for given $S_z$, $\Delta M$\hfil&\cr
height5pt&&\multispan8\hfil&\cr
\noalign{\hrule}
height5pt&&&&&&&&&\omit&\cr
&$9 \hbar$&&26&&20&&7&&1&\cr
height3pt&&&&&&&&&\omit&\cr
&$8 \hbar$&&20&&14&&5&&\omit&\cr
height3pt&&&&&&&&&\omit&\cr
&$7 \hbar$&&14&&11&&3&&\omit&\cr
height3pt&&&&&&&&&\omit&\cr
&$6 \hbar$&&11&&7&&2&&\omit&\cr
height3pt&&&&&&&&&\omit&\cr
&$5 \hbar$&&7&&5&&1&&\omit&\cr
height3pt&&&&&&&&&\omit&\cr
&$4 \hbar$&&5&&3&&1&&\omit&\cr
height3pt&&&&&&&&&\omit&\cr
&$3 \hbar$&&3&&2&&\omit&&\omit&\cr
height3pt&&&&&&&&&\omit&\cr
&$2 \hbar$&&2&&1&&\omit&&\omit&\cr
height3pt&&&&&&&&&\omit&\cr
&$\hbar$&&1&&1&&\omit&&\omit&\cr
height3pt&&&&&&&&&\omit&\cr
&0 (ground)&&1&&\omit&&\omit&&\omit&\cr
height5pt&&&&&&&&&\omit&\cr
\noalign{\hrule}
height5pt&&&&&&&&&\omit&\cr
&$\Delta M$&&$S_z = 0$&&$S_z = \pm1$&&$S_z = \pm2$&&$S_z = \pm3$&\cr
height5pt&&&&&&&&&\omit&\cr}
\hrule\vskip2pt\hrule}\vrule}\hfil
\vfil
\eject
\vfil
\centerline{\bf Table \ref{ttwo}}
\bigskip\bigskip
\hfil\hbox{\vrule\vbox{\offinterlineskip
\hrule\vskip2pt\hrule
\halign {&\phantom{\vrule#}&\quad\hfil#\quad\cr
height2pt&\omit&&\omit&\cr
&19&&0,0,1,1,1,2,2&\cr
height2pt&\omit&&\omit&\cr
&20&&0,0,1,1,2&\cr
height2pt&\omit&&\omit&\cr
&21&&0,1,1&\cr
height2pt&\omit&&\omit&\cr
&22&&0,1&\cr
height2pt&\omit&&\omit&\cr
&23&&1&\cr
height4pt&\omit&&\omit&\cr
&24&&0&\cr
height4pt&\omit&&\omit&\cr
\noalign{\hrule}
height4pt&\omit&&\omit&&\omit&\cr
&$M$&&$S^2$&\cr
height4pt&\omit&&\omit&\cr}
\hrule\vskip2pt\hrule}\vrule}
\hfil
\vfil

\begin{references}

\bibitem[*]{whereamI}{Present address: Department of Physics, MIT,
Cambridge, MA 02139}
\bibitem[\dagger]{email}{E-mail: jemoore@puhep1.princeton.edu}
\bibitem{clark}{R. C. Clark, {\it et al.}, Phys. Rev. Lett. {\bf 60}, 1747 (1988);
Phys. Rev. Lett. {\bf 62}, 1536 (1989).}

\bibitem{eisenstein}{J. P. Eisenstein, H. L. Stormer, L. N. Pfeiffer, and K.W. West,
Phys. Rev. Lett. {\bf 62}, 1540 (1989);
Phys. Rev. B {\bf 41}, 7910 (1990).}

\bibitem{chakra}{T. Chakraborty and F. C. Zhang, Phys. Rev. B. {\bf 29}, 7032 (1984);
E.H. Rezayi, {\it ibid.} {\bf 36}, 5454 (1987).}

\bibitem{maksym}{P. A. Maksym, J. Phys. Condens. Matter {\bf 1}, 6299 (1989).}

\bibitem{jkj}{X. G. Wu, G. Dev, and J. K. Jain, Phys. Rev. Lett {\bf 71}, 153 (1993).}

\bibitem{macdon}{I. A. McDonald, Ph.D. thesis, Princeton University (1994).}

\bibitem{halperin}{B. Halperin,  Helv. Phys. Acta {\bf 56}, 75 (1983).}

\bibitem{laughlin}
{R. B. Laughlin, Phys. Rev. Lett. {\bf 50}, 1395 (1983).}

\bibitem{johnson}{M. D. Johnson and A. H. MacDonald, Phys. Rev. Lett. {\bf 67}
2060, (1991).}

\bibitem{wenrev}{X.-G. Wen, Int. J. Mod. Phys. B {\bf 6}, 1711 (1992).}

\bibitem{wen}{X.-G. Wen, Phys. Rev. B {\bf 41}, 12838 (1990).}

\bibitem{chern}{S. C. Zhang, T. H. Hansson and S. Kivelson, Phys. Rev.
Lett. {\bf 62}, 82 (1989);
N. Read, Phys. Rev. Lett. {\bf 62}, 86 (1989).}

\bibitem{mcd}{I. A. McDonald and F. D. M. Haldane, cond-mat/9511061
(to be published).}

\bibitem{balatsky}{A. Balatsky and M. Stone, Phys. Rev. B {\bf 43}, 8038 (1991).}

\bibitem{hald}{F. D. M. Haldane, lecture in Fractional Statistics and High-$T_c$
Superconductivity Workshop, University of Minnesota.}

\bibitem{pseudo}{F.D.M. Haldane, ''The
Hierarchy of Fractional States and Numerical Studies,'' in
{\sl The Quantum Hall
Effect}, R.E. Prange, S.M. Girvin, eds., Springer-Verlag: New York (1990).}
\end{references}
\end{document}